\documentclass[12pt,twoside]{article}
\usepackage{amssymb}
\usepackage{amsmath}
\usepackage[dvips]{graphicx}
\begin{document}

\begin{center}
{\LARGE  Simulation of two spin-$s$ singlet correlations for all $s$
involving spin measurements}

\vspace{0.6cm} {\bf Ali Ahanj}\footnote{\textsf{Electronic address:
ahanj@physics.unipune.ernet.in}},  {\bf Pramod S.
Joag}\footnote{\textsf{Electronic address:
pramod@physics.unipune.ernet.in}}

\vspace{0.2cm} {\it Department of Physics, University of Pune, Pune
- 411 007, India}

\vspace{0.4cm} and

\vspace{0.2cm} {\bf Sibasish Ghosh}\footnote{\textsf{Electronic
address: sibasish@imsc.res.in}}

\vspace{0.2cm} {\it The Institute of Mathematical Sciences, C. I. T.
Campus, Taramani, Chennai - 600 113, India}
\end{center}

\begin{abstract} In a recent paper [A. Ahanj et al., quant-ph/0603053], we gave a classical
protocol to simulate quantum correlations corresponding to the spin
$s$ singlet state for the infinite sequence of spins satisfying
$2s+1 = 2^{n}$. In the present paper, we have generalized this
result by giving a classical protocol to exactly simulate quantum
correlations implied by the spin-$s$ singlet state corresponding to
all integer as well as half-integer spin values $s$. The class of
measurements we consider here are only those corresponding to spin
observables, as has been done in the above-mentioned paper. The
required amount of communication is found to be $\lceil {\rm
log}_{2} (s + 1) \rceil$ in the worst case scenario, where $\lceil x
\rceil$ is the least integer greater than or equal to $x$.
\end{abstract}

\vspace{0.3cm}
\begin{flushright}
\end{flushright}


\section{Introduction}
It is well known that quantum correlations implied by an entangled
quantum state of a bipartite quantum system cannot be produced
classically, i.e., using only the local and realistic properties of
the subsystems, without any communication between the two subsystems
\cite{bell64}. By quantum correlations we mean the statistical
correlations between the outputs of measurements independently
carried out on each of the two entangled parts. Naturally, the
question arises as to the minimum amount of classical communication
(number of cbits) necessary to simulate the quantum correlations of
an entangled bipartite system. This amount of communication
quantifies the nonlocality of the entangled bipartite quantum
system. It also helps us gauge \cite{methot} the amount of
information hidden in the entangled quantum system itself in some
sense, the amount of information that must be space-like
transmitted, in a local hidden variable model, in order for nature
to account for the excess quantum correlations.

In this scenario, Alice and Bob try and output $\alpha$ and $\beta$
respectively, through a classical protocol, with the same
probability distribution as if they shared the bipartite entangled
system and each measured his or her part of the system according to
a given random Von Neumann measurement. As we have mentioned above,
such a protocol must involve communication between Alice and Bob,
who generally share finite or infinite number of random variables.
The amount of communication is quantified \cite{pironio03} either as
the average number of cbits $\overline {C}(P)$ over the directions
along which the spin components are measured (average or expected
communication) or the worst case communication, which is the maximum
amount of communication $C_{w}(P)$ exchanged between Alice and Bob
in any particular execution of the protocol. The third method is
asymptotic communication i.e., the limit
$lim_{n\rightarrow\infty}\overline{C}(P^{n})$ where $P^{n}$ is the
probability distribution obtained when $n$ runs of the protocol
carried out in parallel i.e., when the parties receive $n$ inputs
and produce $n$ outputs in one go. Note that, naively, Alice can
just tell Bob the direction of her measurement to get an exact
classical simulation, but this corresponds to  an infinite amount of
communication. The question whether a simulation can be done with
finite amount of communication was raised independently by Maudlin
\cite{maudlin92}, Brassard, Cleve and Tapp \cite{brassard99}, and
Steiner \cite{steiner00}. Brassard, Cleve and Tapp used the worst
case communication cost while Steiner used the average. Steiner's
model is weaker as the amount of communication in the worst case can
be unbounded although such cases occur with zero probability.
Brassard, Cleve and Tapp gave a protocol to simulate entanglement in
a singlet state (i.e., the EPR pair) using eight cbits of
communication. Csirik \cite{csirik02} has improved it where one
requires six bits of communication. Toner and Bacon \cite{toner03}
gave a protocol to simulate two-qubit singlet state entanglement
using only one cbit of communication.

Until now, an exact classical simulation of quantum correlations,
for {\it all} possible projective measurements, is accomplished only
for spin $s = 1/2$ singlet state, requiring 1 cbit of classical
communication \cite{toner03}. It is important to know how does the
amount of this classical communication change with the change in the
value of the spin $s$, in order to quantify the advantage offered by
quantum communication over the classical one. Further, this
communication cost quantifies, in terms of classical resources, the
variation of the nonlocal character of quantum correlations with
spin values. In our earlier paper \cite{ali07}, it was shown that
only ${\rm log}_2 (2s + 1)$ bits of communication is needed, in the
worst case scenario, to simulate the measurement correlation of two
spin-$s$ singlet state for performing only measurement of spin
observables on each site, where $s$ is a half-integer spin
satisfying $2s + 1 = 2^n$. Thus these spin values do not include any
integer spin as well as all half-integer spins. In the present
paper we give a classical protocol to simulate the measurement
correlation in a singlet state of two spin-$s$ systems, for all the
integer as well as half-integer values of $s$, considering only (as
above) measurement of spin observables (i.e., measurement of
observables of the form ${\hat{a}}.{\vec{\Lambda}}$ where $\hat{a}$
is any unit vector in ${I\!\!\!R}^3$ and $\vec{\Lambda} =
({\Lambda}_x, {\Lambda}_y, {\Lambda}_z)$ with each ${\Lambda}_i$
being a $(2s + 1) \times (2s + 1)$ traceless Hermitian matrix and
the all three together form the $SU(2)$ algebra). We show that,
using $\lceil {\rm log}_2 (s + 1) \rceil$ bits of classical
communication, one can simulate the above-mentioned measurement
correlation.

We will describe measurement correlations in two spin-$s$ singlet
state in section 2. Before describing our general simulation scheme,
we will explain the scheme with few examples in section 3. In
section 4, we will describe our general simulation scheme. We will
draw our conclusion in section 5.

\section{Singlet state correlation}
The singlet state $|{\psi}^-_s\rangle_{AB}$ of two spin-$s$
particles $A$ and $B$ is the eigenstate corresponding to the
eigenvalue $0$ of the total spin observable of these two spin
systems, namely the state
\begin{equation}
\label{singlet} |{\psi}^-_s\rangle_{AB} = \frac{1}{\sqrt{2s + 1}} \sum_{m =
-s}^{s} (- 1)^{s - m} |m\rangle_A \otimes |- m\rangle_B,
\end{equation}
where $|- s\rangle$, $|- s + 1\rangle$, $\ldots$, $|s - 1\rangle$,
$|s\rangle$ are eigenstates of the spin observable of each of the
individual spin-$s$ system. Thus $|{\psi}^-_s\rangle_{AB}$ is a
maximally entangled state of the bipartite system $A + B$, described
by the Hilbert space ${C\!\!\!\!I}^{2s + 1} \otimes {C\!\!\!\!I}^{2s
+ 1}$.

We will consider here measurement of `spin observables', namely the
observables of the form $\hat{a}.{\bf J}$ on each individual
spin-$s$ system, where $\hat{a}$ is an arbitrary unit vector in
${I\!\!\!R}^3$ and ${\bf J} = (J_x, J_y, J_z)$ (see ref.
\cite{ali07} for a discussion on the choice of measurement
observables). For the $(2s + 1) \times (2s + 1)$ matrix
representations of the spin observables $J_x$, $J_y$, and $J_z$,
please see  page 191 - 192 of ref. \cite{sakurai99}. $J$ matrices
satisfy the $SU(2)$ algebra, namely $[J_x, J_y] = iJ_z$, $[J_y, J_z]
= iJ_x$, $[J_z, J_x] = iJ_y$. The eigenvalues of $\hat{a}.{\bf J}$
are $- s$, $- s + 1$, $\ldots$, $s - 1$, $s$ for all $\hat{a} \in
{I\!\!\!R}^3$. The quantum correlations
$\langle{\psi}^-_s|\hat{a}.{\bf J} \otimes \hat{b}.{\bf
J}|{\psi}^-_s\rangle$ (which we will denote here as $\left\langle
\alpha \beta\right\rangle$, where $\alpha$ runs through all the
eigenvalues of $\hat{a}.{\bf J}$ and $\beta$ runs through all the
eigenvalues of $\hat{b}.{\bf J}$) is given by
\begin{equation}
\label{correlation} \langle{\psi}^-_s|\hat{a}.{\bf J} \otimes
\hat{b}.{\bf J}|{\psi}^-_s\rangle = \left\langle \alpha
\beta\right\rangle=-\frac{1}{3} s(s+1)\hat{a}.\hat{b} \;,
\end{equation}
where $\hat{a}$ and $\hat{b}$ are the unit vectors specifying the
directions along which the spin components are measured by Alice and
Bob respectively (see section 6-6 of page 179 in \cite{peres93}).
Note that, by virtue of being a singlet state ,$\left\langle \alpha
\right\rangle = 0 = \left\langle\beta\right\rangle$ irrespective of
directions $\hat{a} $ and $\hat{b}$.

Let us now come to our protocol. In the simulation of the
measurement of the observable $\hat{a}.{\bf J}$ (where $\hat{a} \in
{I\!\!\!R}^3$ is the supplied direction of measurement), Alice will
have to reproduce the $2s + 1$ number of outcomes $\alpha = s, s -
1, \ldots, - s + 1, - s$ with equal probability. Similarly, Bob will
have to reproduce the $2s + 1$ number of outcomes $\beta = s, s - 1,
\ldots, - s + 1, - s$ with equal probability. We will describe our
protocol for the simulation by first giving the ones for smaller
values of the spin and then by giving the protocol for general value
of the spin.

Before describing the simulation scheme, we mention here few
mathematical results which will be frequently needed during our
discussion of the simulation scheme. Consider the unit sphere in
three dimensional Euclidean space: $S_2 = \{|{\bf r}| = 1 : {\bf r}
\in {I\!\!\!R}^3\}$. Let ${\hat{\lambda}}_1$, ${\hat{\lambda}}_2$,
${\hat{\mu}}_1$, ${\hat{\mu}}_2$, ${\hat{\nu}}_1$, ${\hat{\nu}}_2$
be (mutually) independent but uniformly distributed random variables
on $S_2$. Let $\hat{a}$ and $\hat{b}$ be given any two elements from
$S_2$. Also $\hat{z}$ be the unit vector along the $z$-axis of the
rectangular Cartesian co-ordinate axes $x$, $y$ and $z$ -- the
associated reference frame. Let us define:
$$c_k = {\rm Sgn} (\hat{a}.{\hat{\lambda}}_k)~ {\rm Sgn}
(\hat{a}.{\hat{\mu}}_k)~~ (k = 1, 2),$$
$$f_k = {\rm Sgn} \left(\hat{z}.{\hat{\nu}}_k + p_k\right)~~ (p_k \in
(0, 1)),$$ where ${\rm Sgn} : {I\!\!\!R} \rightarrow \{+1, -1\}$ is
the function defined as ${\rm Sgn} (x) = 1$ if $x \ge 0$ and ${\rm
Sgn} (x) = -1$ if $x < 0$. One can show that (see ref.
\cite{toner03} for the derivations):
\begin{equation}
\label{alphadescription} {\rm Prob} \left({\rm Sgn}
\left(\hat{a}.{\hat{\lambda}}_k\right) = {\pm}1\right) =
\frac{1}{2},~~ ({\rm for}~ k = 1, 2),
\end{equation}
and hence
\begin{equation}
\label{alphades1} \left\langle {\rm Sgn}
\left(\hat{a}.{\hat{\lambda}}_k\right) \right\rangle = 0~~ ({\rm
for}~ k = 1, 2).
\end{equation}
\begin{equation}
\label{betadescription} {\rm Prob} \left({\rm Sgn}
\left[\hat{b}.\left({\hat{\lambda}}_k +
c_k{\hat{\mu}}_k\right)\right] = {\pm}1\right) = \frac{1}{2},~~
({\rm for}~ k = 1, 2),
\end{equation}
and hence
\begin{equation}
\label{betades1} \left\langle {\rm Sgn}
\left[\hat{b}.\left({\hat{\lambda}}_k +
c_k{\hat{\mu}}_k\right)\right] \right\rangle = 0~~ ({\rm for}~ k =
1, 2).
\end{equation}
\begin{equation}
\label{alphabeta} \left\langle {\rm Sgn}
\left(\hat{a}.{\hat{\lambda}}_k\right) \times {\rm Sgn}
\left[\hat{b}.\left({\hat{\lambda}}_l +
c_l{\hat{\mu}}_l\right)\right] \right\rangle =
{\delta}_{kl}\left(\hat{a}.\hat{b}\right)~~ ({\rm for}~ k, l = 1,
2).
\end{equation}
 Also we have (taking ${\hat{\nu}}_k = ({\rm sin}
{\theta}_k~ {\rm cos} {\phi}_k,~ {\rm sin} {\theta}_k~ {\rm sin}
{\phi}_k,~ {\rm cos} {\theta}_k)$)
\begin{equation}
\label{f+} {\rm Prob} \left(f_k = +1\right) =
\frac{1}{4\pi}\int_{{\phi}_k = 0}^{{\phi}_k = 2\pi} \int_{{\theta}_k
= 0}^{{\theta}_k = {\rm cos}^{-1} (-p_k)} {\rm sin} {\theta}_k
d{\theta}_k d{\phi}_k = \frac{1 + p_k}{2}~~ ({\rm for}~ k = 1, 2),
\end{equation}
and hence
\begin{equation}
\label{f-} {\rm Prob} \left(f_k = -1\right) = \frac{1 - p_k}{2}~~
({\rm for}~ k = 1, 2).
\end{equation}
So \begin{equation} \label{faverage} \left\langle f_k \right\rangle
= p_k~~ ({\rm for}~ k = 1, 2).
\end{equation}
Moreover, as $f_k^2$ will always have the value $+1$, therefore
$$\left\langle f_k^2 \right\rangle = 1~~ ({\rm for}~ k = 1, 2).$$
Consequently
\begin{equation}
\label{faveragenew} \left\langle \left(1 + f_k\right)^2
\right\rangle = 2\left(1 + p_k\right)~~ ({\rm for}~ k = 1, 2)
\end{equation}
and (as ${\hat{\nu}}_1$ and ${\hat{\nu}}_2$ are independent random
variables)
\begin{equation}
\label{f2average} \left\langle \left(1 + f_1\right)^2\left(1 +
f_2\right)^2 \right\rangle = \left\langle \left(1 + f_1\right)^2
\right\rangle \times \left\langle \left(1 + f_2\right)^2
\right\rangle = 4\left(1 + p_1\right)\left(1 + p_2\right).
\end{equation}
Again, as ${\hat{\lambda}}_1$, ${\hat{\lambda}}_2$, ${\hat{\mu}}_1$,
${\hat{\mu}}_2$, ${\hat{\nu}}_1$, ${\hat{\nu}}_2$ are independent
random variables, therefore
\begin{equation}
\label{genaverage} \left\langle \left(1 + f_k\right)^2 \times {\rm
Sgn} \left(\hat{a}.{\hat{\lambda}}_l\right) \times {\rm Sgn}
\left[\hat{b}.\left({\hat{\lambda}}_m +
c_l{\hat{\mu}}_m\right)\right] \right\rangle = 2\left(1 +
p_k\right){\delta}_{lm}\left(\hat{a}.\hat{b}\right),
\end{equation}
and
\begin{equation}
\label{gen2average} \left\langle \left(1 + f_1\right)^2\left(1 +
f_2\right)^2 \times {\rm Sgn} \left(\hat{a}.{\hat{\lambda}}_k\right)
\times {\rm Sgn} \left[\hat{b}.\left({\hat{\lambda}}_l +
c_l{\hat{\mu}}_l\right)\right] \right\rangle = 4\left(1 +
p_1\right)\left(1 +
p_2\right){\delta}_{kl}\left(\hat{a}.\hat{b}\right).
\end{equation}

\section{Examples}
For each value $s$ of the spin, we can always find a positive
integer $n$ such that $2^{n - 1} < s + 1 \le 2^n$. We show here
below that the above-mentioned simulation can be done with just $n$
bits of communication if $s$ is such that $2^{n - 1} < s + 1 \le
2^n$. To give a clear picture, let us first describe our protocol
for few lower values of $s$, and after that, the general protocol
will be given. To start with, Alice and Bob fix a common reference
frame (with rectangular Cartesian co-ordinate axes $x$, $y$ and $z$)
for them.

\vspace{0.2cm} {\noindent {\bf Example 1:} $2^{1 - 1} < s + 1 \le
2^1$. Thus the allowed values of $s$ are $1/2$ and $1$.}

\vspace{0.2cm} {\underline{{\bf Case (1.1)} $s = 1/2$:}

Alice and Bob {\it a priori} share two independent and uniformly
distributed random variables ${\hat{\lambda}}_{1/2}$,
${\hat{\mu}}_{1/2} \in S_2$. Given the measurement direction
$\hat{a} \in S_2$, Alice calculates her output as $\alpha = -(1/2)
{\rm Sgn} (\hat{a}.{\hat{\lambda}}_{1/2}) \equiv -\alpha(1/2)$
(say). She also sends the bit value $c_{1/2} = {\rm Sgn}
(\hat{a}.{\hat{\lambda}}_{1/2})~ {\rm Sgn} (\hat{a}.
{\hat{\mu}}_{1/2})$ to Bob by classical communication. After
receiving this bit value and using the supplied measurement
direction $\hat{b} \in S_2$, Bob now calculates his output as $\beta
= (1/2) {\rm Sgn} [\hat{b}.({\hat{\lambda}}_{1/2} +
c_{1/2}{\hat{\mu}}_{1/2})] \equiv \beta(1/2)$ (say). It is known
that (see equations (\ref{alphades1}) - (\ref{alphabeta})) for the
two spin-$1/2$ singlet state $|{\psi}^-_{1/2}\rangle$, $\alpha,
\beta \in \{+1/2, -1/2\}$, ${\rm Prob} (\alpha = {\pm}1/2) = {\rm
Prob} (\beta = {\pm}1/2) = 1/2$ (and so $\langle \alpha \rangle =
\langle \beta \rangle = 0$), and $\langle \alpha \beta \rangle =
-(1/3)(1/2)(1/2 + 1)\hat{a}.\hat{b} = \langle \alpha \beta
\rangle_{QM}$. Thus the total number of cbits required (we denote it
by $n_c$), for simulating the measurement correlation in the worst
case scenario, is one and the total number of shared random variable
is two: ${\lambda}_{1/2}$ and ${\mu}_{1/2}$. Thus here $n_{\lambda}
\equiv$ the total number of $\hat{\lambda}$'s $= 1$ and $n_{\mu}
\equiv$ the total number of $\hat{\mu}$'s $= 1$.

\vspace{0.2cm} {\underline{{\bf Case (1.2)} $s = 1$:}

Alice and Bob {\it a priori} share three independent and uniformly
distributed random variables ${\hat{\lambda}}_{1}$,
${\hat{\mu}}_{1}$, ${\hat{\nu}}_{1} \in S_2$. Given the measurement
direction $\hat{a} \in S_2$, Alice calculates her output as $\alpha
= -((1 + f_1)/2) {\rm Sgn} (\hat{a}.{\hat{\lambda}}_{1}) \equiv
-\alpha(1)$ (say). She also sends the bit value $c_{1} = {\rm Sgn}
(\hat{a}.{\hat{\lambda}}_{1})~ {\rm Sgn} (\hat{a}. {\hat{\mu}}_{1})$
to Bob by classical communication. After receiving this bit value
and using the supplied measurement direction $\hat{b} \in S_2$, Bob
now calculates his output as $\beta = ((1 + f_1)/2) {\rm Sgn}
[\hat{b}.({\hat{\lambda}}_{1} + c_{1}{\hat{\mu}}_{1})] \equiv
\beta(1)$ (say), where $f_1 = {\rm Sgn} (\hat{z}.{\hat{\nu}}_{1} +
1/3)$ and $c_1 = {\rm Sgn} (\hat{a}.{\hat{\lambda}}_1)~ {\rm Sgn}
(\hat{a}.{\hat{\mu}}_1)$. Now, by equations (\ref{f+}) -
(\ref{faverage}), we have ${\rm Prob} (f_1 = + 1) = 2/3$, ${\rm
Prob} (f_1 = - 1) = 1/3$ and $\langle f_1 \rangle = 1/3$. Thus we
see that (using equations (\ref{alphadescription}),
(\ref{betadescription}), the probability distribution of $f_1$, and
the fact that ${\hat{\lambda}}_1$, ${\hat{\mu}}_1$, ${\hat{\nu}}_1$
are independent random variables) $\alpha, \beta \in \{+1, 0, -1\}$
and ${\rm Prob} (\alpha = j) = {\rm Prob} (\beta = k) = 1/3$ for all
$j, k \in \{+1, 0, -1\}$. Also we have (using equation
(\ref{genaverage})) $\langle \alpha \beta \rangle = -(1/3) \times 1
\times (1 + 1)\hat{a}.\hat{b} = \langle \alpha \beta \rangle_{QM}$.
Thus here $n_c = 1$, $n_{\lambda} = 1$, $n_{\mu} = 1$, $n_{\nu}
\equiv$ the total number of $\hat{\nu}$'s $= 1$.

\vspace{0.4cm} {\noindent {\bf Example 2:} $2^{2 - 1} < s + 1 \le
2^2$. Thus the allowed values of $s$ are $3/2$, $2$, $5/2$, and
$3$.}

\vspace{0.2cm} {\underline{{\bf Case (2.1)} $s = 3/2$:}

Alice and Bob {\it a priori} share four independent and uniformly
distributed random variables ${\hat{\lambda}}_{1/2}$,
${\hat{\lambda}}_{3/2}$, ${\hat{\mu}}_{1/2}$, ${\hat{\mu}}_{3/2} \in
S_2$. Given the measurement direction $\hat{a} \in S_2$, Alice
calculates her output as $\alpha = - [{\rm Sgn}
(\hat{a}.{\hat{\lambda}}_{3/2}) + \alpha(1/2)] \equiv -\alpha(3/2)$
(say), where $\alpha(1/2)$ involves ${\hat{\lambda}}_{1/2}$ and is
described in (1.1) above. She also sends the two bit values $c_{k} =
{\rm Sgn} (\hat{a}.{\hat{\lambda}}_{k})~ {\rm Sgn} (\hat{a}.
{\hat{\mu}}_{k})$ (for $k = 1/2, 3/2$) to Bob by classical
communication. After receiving these two bit values and using the
supplied measurement direction $\hat{b} \in S_2$, Bob now calculates
his output as $\beta = {\rm Sgn} [\hat{b}.({\hat{\lambda}}_{3/2} +
c_{3/2}{\hat{\mu}}_{3/2})] + \beta(1/2) \equiv \beta(3/2)$ (say),
where $\beta(1/2)$ involves ${\hat{\lambda}}_{1/2}$,
${\hat{\mu}}_{1/2}$ and is described in (1.1) above. Using equations
(\ref{alphadescription}) and (\ref{betadescription}), and using the
fact that ${\hat{\lambda}}_{1/2}$, ${\hat{\lambda}}_{3/2}$,
${\hat{\mu}}_{1/2}$, ${\hat{\mu}}_{3/2}$ are independent and
uniformly distributed random variables on $S_2$, we have ${\rm Prob}
(\alpha = j) = {\rm Prob} (\beta = k) = 1/4$ for all $j, k \in
\{+3/2, +1/2, -1/2, -3/2\}$. Also, by using equation
(\ref{alphabeta}), we have $\langle \alpha \beta \rangle = -(1/3)
\times (3/2) \times (3/2 + 1)\hat{a}.\hat{b} = \langle \alpha \beta
\rangle_{QM}$. Thus here $n_c = 2$, $n_{\lambda} = 2$, $n_{\mu} = 2$
and $n_{\nu} = 0$.

\vspace{0.2cm} {\underline{{\bf Case (2.2)} $s = 2$:}

Alice and Bob {\it a priori} share five independent and uniformly
distributed random variables ${\hat{\lambda}}_{1/2}$,
${\hat{\lambda}}_{2}$, ${\hat{\mu}}_{1/2}$, ${\hat{\mu}}_{2}$,
${\hat{\nu}}_{2} \in S_2$. Given the measurement direction $\hat{a}
\in S_2$, Alice calculates her output as $\alpha = - ((1 +
f_2)/2)[(3/2){\rm Sgn} (\hat{a}.{\hat{\lambda}}_{2}) + \alpha(1/2)]
\equiv -\alpha(2)$ (say), where $\alpha(1/2)$ involves
${\hat{\lambda}}_{1/2}$ and is described in (1.1) above. She also
sends the two bit values $c_{k} = {\rm Sgn}
(\hat{a}.{\hat{\lambda}}_{k})~ {\rm Sgn} (\hat{a}. {\hat{\mu}}_{k})$
(for $k = 1/2, 2$) to Bob by classical communication. After
receiving these two bit values and using the supplied measurement
direction $\hat{b} \in S_2$, Bob now calculates his output as $\beta
= ((1 + f_2)/2)[(3/2){\rm Sgn} [\hat{b}.({\hat{\lambda}}_{2} +
c_{2}{\hat{\mu}}_{2})] + \beta(1/2)] \equiv \beta(2)$ (say),
where $\beta(1/2)$ involves ${\hat{\lambda}}_{1/2}$,
${\hat{\mu}}_{1/2}$ and is described in (1.1) above. Here $f_2 =
{\rm Sgn} (\hat{z}.{\hat{\nu}}_{2} + 3/5)$. By using equations
(\ref{f+}) - (\ref{faverage}), we see that ${\rm Prob} (f_2 = +1) =
4/5$, ${\rm Prob} (f_2 = -1) = 1/5$ and $\langle f_2 \rangle = 3/5$.
Using these facts and the fact that ${\hat{\lambda}}_{1/2}$,
${\hat{\lambda}}_{2}$, ${\hat{\mu}}_{1/2}$, ${\hat{\mu}}_{2}$,
${\hat{\nu}}_{2}$ are independent and uniformly distributed random
variables on $S_2$, we have ${\rm Prob} (\alpha = j) = {\rm Prob}
(\beta = k) = 1/5$ for all $j, k \in \{+2, +1, 0, -1, -2\}$. Also,
by using equation (\ref{genaverage}) $\langle \alpha \beta \rangle =
-(1/3) \times 2 \times (2 + 1)\hat{a}.\hat{b} = \langle \alpha \beta
\rangle_{QM}$. Thus here $n_c = 2$, $n_{\lambda} = 2$, $n_{\mu} = 2$
and $n_{\nu} = 1$.

\vspace{0.2cm} {\underline{{\bf Case (2.3)} $s = 5/2$:}

Alice and Bob {\it a priori} share five independent and uniformly
distributed random variables ${\hat{\lambda}}_{1}$,
${\hat{\lambda}}_{5/2}$, ${\hat{\mu}}_{1}$, ${\hat{\mu}}_{5/2}$,
${\hat{\nu}}_{1} \in S_2$. Given the measurement direction $\hat{a}
\in S_2$, Alice calculates her output as $\alpha = - [(3/2){\rm Sgn}
(\hat{a}.{\hat{\lambda}}_{5/2}) + \alpha(1)] \equiv -\alpha(5/2)$
(say), where $\alpha(1)$ involves ${\hat{\lambda}}_{1}$,
${\hat{\nu}}_1$ and is described in (1.2) above. She also sends the
two bit values $c_{k} = {\rm Sgn} (\hat{a}.{\hat{\lambda}}_{k})~
{\rm Sgn} (\hat{a}. {\hat{\mu}}_{k})$ (for $k = 1, 5/2$) to Bob by
classical communication. After receiving these two bit values and
using the supplied measurement direction $\hat{b} \in S_2$, Bob now
calculates his output as $\beta = (3/2){\rm Sgn}
[\hat{b}.({\hat{\lambda}}_{5/2} + c_{5/2}{\hat{\mu}}_{5/2})] +
\beta(1) \equiv \beta(5/2)$ (say), where $\beta(1)$ involves
${\hat{\lambda}}_{1}$, ${\hat{\mu}}_{1}$, ${\hat{\nu}}_1$ and is
described in (1.2) above. Using the fact that ${\hat{\lambda}}_{1}$,
${\hat{\lambda}}_{5/2}$, ${\hat{\mu}}_{1}$, ${\hat{\mu}}_{5/2}$,
${\hat{\nu}}_{1}$ are independent and uniformly distributed random
variables on $S_2$, equations (\ref{alphadescription}) and
(\ref{betadescription}), and the discussions in (1.2) above, we have
${\rm Prob} (\alpha = j) = {\rm Prob} (\beta = k) = 1/6$ for all $j,
k \in \{+5/2, +3/2, +1/2, -1/2, -3/2, -5/2\}$. Also, by using
equation (\ref{genaverage}) $\langle \alpha \beta \rangle = -(1/3)
\times (5/2) \times (5/2 + 1)\hat{a}.\hat{b} = \langle \alpha \beta
\rangle_{QM}$. Thus here $n_c = 2$, $n_{\lambda} = 2$, $n_{\mu} = 2$
and $n_{\nu} = 1$.

\vspace{0.2cm} {\underline{{\bf Case (2.4)} $s = 3$:}

Alice and Bob {\it a priori} share six independent and uniformly
distributed random variables ${\hat{\lambda}}_{1}$,
${\hat{\lambda}}_{3}$, ${\hat{\mu}}_{1}$, ${\hat{\mu}}_{3}$,
${\hat{\nu}}_{1}$, ${\hat{\nu}}_3 \in S_2$. Given the measurement
direction $\hat{a} \in S_2$, Alice calculates her output as $\alpha
= - ((1 + f_3)/2)[2{\rm Sgn} (\hat{a}.{\hat{\lambda}}_{3}) +
\alpha(1)] \equiv -\alpha(3)$ (say), where $\alpha(1)$ involves
${\hat{\lambda}}_{1}$, ${\hat{\nu}}_1$ and is described in (1.2)
above. Here $f_3 = {\rm Sgn} (\hat{z}.{\hat{\nu}}_3 + 5/7)$. She
also sends the two bit values $c_{k} = {\rm Sgn}
(\hat{a}.{\hat{\lambda}}_{k})~ {\rm Sgn} (\hat{a}. {\hat{\mu}}_{k})$
(for $k = 1, 3$) to Bob by classical communication. After receiving
these two bit values and using the supplied measurement direction
$\hat{b} \in S_2$, Bob now calculates his output as $\beta = ((1 +
f_3)/2)[2{\rm Sgn} [\hat{b}.({\hat{\lambda}}_{3} +
c_{3}{\hat{\mu}}_{3})] + \beta(1)] \equiv \beta(3)$ (say), where
$\beta(1)$ involves ${\hat{\lambda}}_{1}$, ${\hat{\mu}}_{1}$,
${\hat{\nu}}_1$ and is described in (1.2) above. Using the fact that
${\hat{\lambda}}_{1}$, ${\hat{\lambda}}_{3}$, ${\hat{\mu}}_{1}$,
${\hat{\mu}}_{3}$, ${\hat{\nu}}_{1}$, ${\hat{\nu}}_3$ are
independent and uniformly distributed random variables on $S_2$,
equations (\ref{alphadescription}) and (\ref{betadescription}), and
the discussions in (1.2) above, we have ${\rm Prob} (\alpha = j) =
{\rm Prob} (\beta = k) = 1/7$ for all $j, k \in \{+3, +2, +1, 0, -1,
-2, -3\}$. Also, by using equations (\ref{genaverage}) and
(\ref{gen2average}), we have $\langle \alpha \beta \rangle = -(1/3)
\times 3 \times (3 + 1)\hat{a}.\hat{b} = \langle \alpha \beta
\rangle_{QM}$. Thus here $n_c = 2$, $n_{\lambda} = 2$, $n_{\mu} = 2$
and $n_{\nu} = 2$.

\section{General simulation scheme}
Let us now describe the protocol for general $s$. One can always
find out uniquely a positive integer $n$ such that $2^{n - 1} < s +
1 \le 2^n$. Equivalently, given the dimension $d = 2s + 1$ of the
Hilbert space, one can always find out a unique positive integer $n$
such that $2^{n} - 1 < d \le 2^{n + 1} - 1$. Let $d = a_02^n +
a_12^{n - 1} + \ldots + a_n2^0 \equiv \underline{a_0a_1 \ldots a_n}$
be the binary representation of $d$ (where $a_0$, $a_1$, $\ldots$,
$a_n \in \{0, 1\}$). So we must have $a_0 \ne 0$. Before describing
the general simulation scheme, using the help of the above-mentioned
examples, let us describe below the scheme pictorially (see Figure
1) in terms of binary representation of the dimension of the
individual spin system. The simulation scheme, we have described in
ref. \cite{ali07} for the simulation of the measurement correlation
in two spin-$s$ singlet state, where $2s + 1 = 2^n$, corresponds to
the upper most chain
$$2^1 = \underline{10} \rightarrow 2^2 = \underline{100} \rightarrow
2^3 = \underline{1000} \rightarrow \ldots \rightarrow 2^{n - 1} =
\underline{1000 \ldots 00} \rightarrow 2^n = \underline{1000 \ldots
000}$$ in Figure 1. In other words, when $2s + 1 = 2^n$, given the
measurement directions $\hat{a}$, Alice will calculate her output
$-\alpha\left(\frac{2^n - 1}{2}\right) \equiv
-\alpha\left(\frac{\underline{1000 \ldots 000} - 1}{2}\right)$ as:
$$-\alpha\left(\frac{\underline{1000 \ldots 000} - 1}{2}\right) =
-\left[\left(\frac{\frac{\underline{1000 \ldots 000} - 1}{2} +
\frac{1}{2}}{2}\right) {\rm Sgn}
\left(\hat{a}.{\hat{\lambda}}_{\frac{\underline{1000 \ldots 000} -
1}{2}}\right) + \alpha\left(\frac{\underline{1000 \ldots 00} -
1}{2}\right)\right]$$
$$= -[\left(\frac{\frac{\underline{1000 \ldots 000} - 1}{2} +
\frac{1}{2}}{2}\right) {\rm Sgn}
\left(\hat{a}.{\hat{\lambda}}_{\frac{\underline{1000 \ldots 000} -
1}{2}}\right) + \left(\frac{\frac{\underline{1000 \ldots 00} - 1}{2}
+ \frac{1}{2}}{2}\right) {\rm Sgn}
\left(\hat{a}.{\hat{\lambda}}_{\frac{\underline{1000 \ldots 00} -
1}{2}}\right) +$$
$$\alpha\left(\frac{\underline{1000 \ldots 0}
- 1}{2}\right)]$$
$$\ldots$$
$$\ldots$$
$$= -[\left(\frac{\frac{\underline{1000 \ldots 000} - 1}{2} +
\frac{1}{2}}{2}\right) {\rm Sgn}
\left(\hat{a}.{\hat{\lambda}}_{\frac{\underline{1000 \ldots 000} -
1}{2}}\right) + \left(\frac{\frac{\underline{1000 \ldots 00} - 1}{2}
+ \frac{1}{2}}{2}\right) {\rm Sgn}
\left(\hat{a}.{\hat{\lambda}}_{\frac{\underline{1000 \ldots 00} -
1}{2}}\right) +$$
$$\ldots + \left(\frac{\frac{\underline{10} -
1}{2} + \frac{1}{2}}{2}\right) {\rm Sgn}
\left(\hat{a}.{\hat{\lambda}}_{\frac{\underline{10} -
1}{2}}\right)]$$
$$= -\frac{1}{2}\sum_{k = 1}^{n} 2^{n - k} {\rm Sgn} \left(\hat{a}.{\hat{\eta}}_k\right),$$
where ${\hat{\eta}}_k = {\hat{\lambda}}_{\frac{2^k - 1}{2}}$.
Similarly for Bob. We have generalized below this scheme to
arbitrary value of $s$ (see equations (\ref{halfintegeralpha}) -
(\ref{integerbeta})).

\begin{figure}

\begin{center}
\includegraphics[width=14cm,height=18cm]{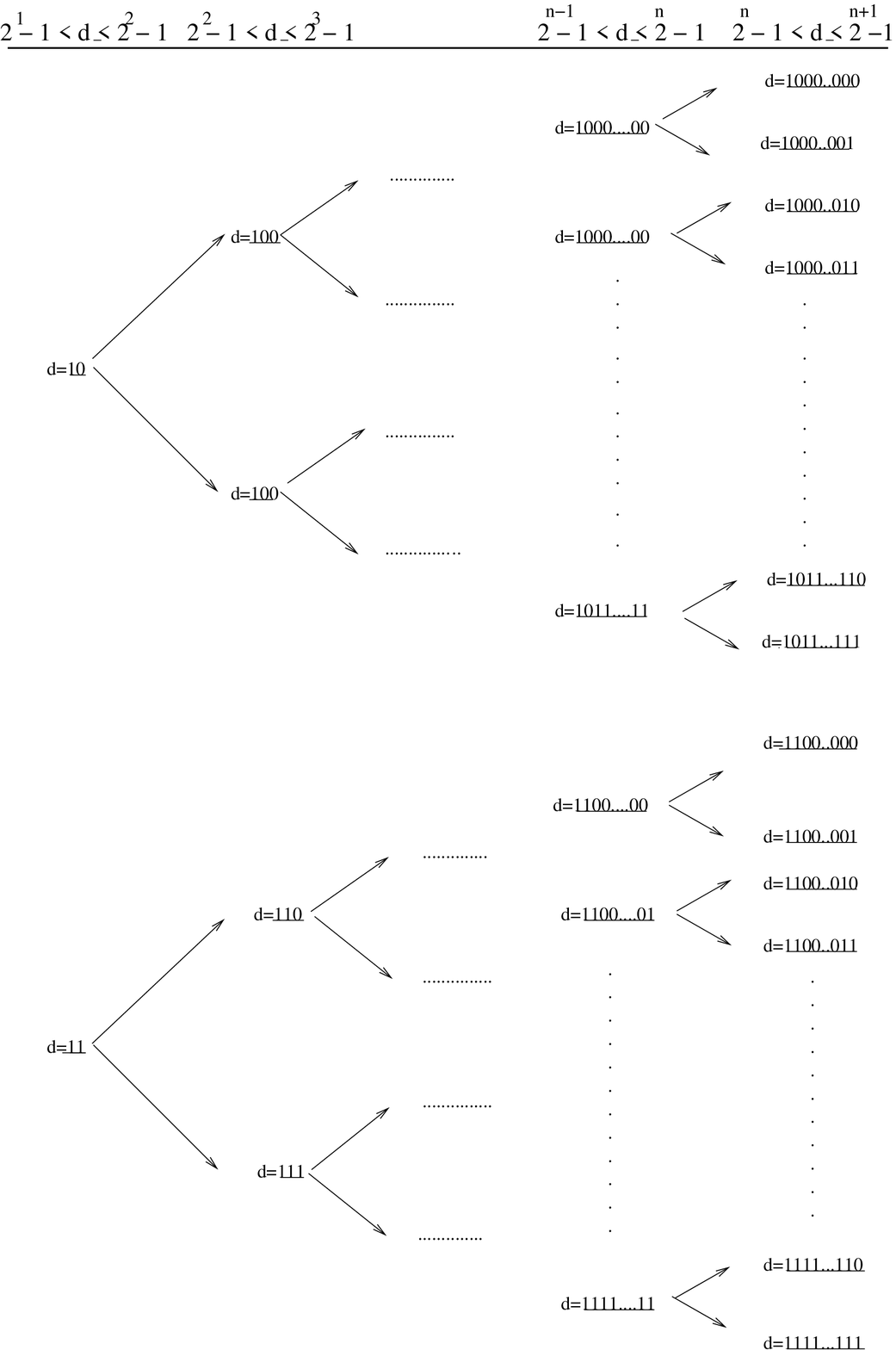}

\vspace{0.4cm}
\underline{Figure 1:}\small{ The paths (mentioned by
concatenated arrows from left to right) of simulation for each
integer and half-integer spins $s$ such that $2^{n - 1} < s + 1 \le
2^n$}
\end{center}

\end{figure}

\newpage

To describe the general simulation, we consider the following two
cases:

\vspace{0.2cm} {\noindent \underline{\bf $s$ is a half-integer
spin:}}}

Over and above the $n - 1$ number of ${\hat{\lambda}}$'s, $n - 1$
number of ${\hat{\mu}}$'s and $(a_1 + a_2 + \ldots + a_{n - 1})$
number of ${\hat{\nu}}$'s appeared in the expression for
$\alpha\left(\frac{\underline{a_0a_1 \ldots a_{n - 1}} -
1}{2}\right)$ and $\beta\left(\frac{\underline{a_0a_1 \ldots a_{n -
1}} - 1}{2}\right)$, Alice and Bob share the random variables
${\hat{\lambda}}_{\frac{\underline{a_0a_1 \ldots a_n} - 1}{2}}$ and
${\hat{\mu}}_{\frac{\underline{a_0a_1 \ldots a_n} - 1}{2}}$, where,
it has been assumed that all these $2n + (a_1 + a_2 + \ldots a_{n -
1})$ number of random variables are independent and uniformly
distributed on $S_2$. Let us denote the set of all these $n$
${\hat{\lambda}}$'s by $S_{\lambda}$, the set of all these $n$
${\hat{\mu}}$'s by $S_{\mu}$, and the set of all these $(a_1 + a_2 +
\ldots + a_{n - 1})$ ${\hat{\nu}}$'s by $S_{\nu}$. Given the
measurement direction $\hat{a} \in S_2$, Alice calculates her output
as
 $$\alpha =
-\left[\left(\frac{\frac{\underline{a_0a_1 \ldots a_n} - 1}{2} +
\frac{1}{2}}{2}\right) {\rm Sgn}
\left(\hat{a}.{\hat{\lambda}}_{\frac{\underline{a_0a_1 \ldots a_n} -
1}{2}}\right) + \alpha\left(\frac{\underline{a_0a_1 \ldots a_{n -
1}} - 1}{2}\right)\right]$$
\begin{equation}
\label{halfintegeralpha}
\equiv -\alpha\left(\frac{\underline{a_0a_1
\ldots a_n} - 1}{2}\right),
\end{equation} and she sends the $n$ cbits
\begin{equation}
\label{cbitshalf} c_{k} = {\rm Sgn}
(\hat{a}.{\hat{\lambda}}_k)~ {\rm Sgn} (\hat{a}.{\hat{\mu}}_k),
\end{equation}
to Bob where $k = \frac{\underline{a_0a_1 \ldots a_n} - 1}{2},
\frac{\underline{a_0a_1 \ldots a_{n - 1}} - 1}{2}, \ldots,
\frac{\underline{a_0a_1} - 1}{2}$. After receiving these $n$ cbits
and using his measurement direction $\hat{b} \in S_2$, Bob
calculates his output as
$$\beta =
\left[\left(\frac{\frac{\underline{a_0a_1 \ldots a_n} - 1}{2} +
\frac{1}{2}}{2}\right) {\rm Sgn}
\left[\hat{b}.\left({\hat{\lambda}}_{\frac{\underline{a_0a_1 \ldots
a_n} - 1}{2}} + c_{\frac{\underline{a_0a_1 \ldots a_n} -
1}{2}}{\hat{\mu}}_{\frac{\underline{a_0a_1 \ldots a_n} -
1}{2}}\right)\right]\right.$$
\begin{equation}
\label{halfintegerbeta} \left.+ \beta\left(\frac{\underline{a_0a_1
\ldots a_{n - 1}} - 1}{2}\right)\right] \equiv
\beta\left(\frac{\underline{a_0a_1 \ldots a_n} - 1}{2}\right).
\end{equation}
Let $L = a_1 + a_2 + \ldots a_n$ and let $i_1$, $i_2$, $\ldots$,
$i_L$ be all those elements from $\{1, 2, \ldots, n\}$ such that
$i_1 < i_2 < \ldots < i_L$ and $a_{i_1} = a_{i_2} = \ldots = a_{i_L}
= 1$. It is then easy to see that
$$S_{\lambda} = \left\{{\hat{\lambda}}_{\frac{\underline{a_0a_1} -
1}{2}}, {\hat{\lambda}}_{\frac{\underline{a_0a_1a_2} - 1}{2}},
\ldots, {\hat{\lambda}}_{\frac{\underline{a_0a_1 \ldots a_n} -
1}{2}}\right\},$$
$$S_{\mu} = \left\{{\hat{\mu}}_{\frac{\underline{a_0a_1} -
1}{2}}, {\hat{\mu}}_{\frac{\underline{a_0a_1a_2} - 1}{2}}, \ldots,
{\hat{\mu}}_{\frac{\underline{a_0a_1 \ldots a_n} - 1}{2}}\right\},$$
$$S_{\nu} = \left\{{\hat{\nu}}_{\frac{\underline{a_0a_{i_1}} -
1}{2}}, {\hat{\nu}}_{\frac{\underline{a_0a_{i_1}a_{i_2}} - 1}{2}},
\ldots, {\hat{\nu}}_{\frac{\underline{a_0a_{i_1} \ldots a_{i_L}} -
1}{2}}\right\}.$$

\vspace{0.2cm} {\noindent \underline{\bf $s$ is an integer spin:}}}

Over and above the $n - 1$ number of ${\hat{\lambda}}$'s, $n - 1$
number of ${\hat{\mu}}$'s and $(a_1 + a_2 + \ldots + a_{n - 1})$
number of ${\hat{\nu}}$'s appeared in the expression for
$\alpha\left(\frac{\underline{a_0a_1 \ldots a_{n - 1}} -
1}{2}\right)$ and $\beta\left(\frac{\underline{a_0a_1 \ldots a_{n -
1}} - 1}{2}\right)$, Alice and Bob share the random variables
${\hat{\lambda}}_{\frac{\underline{a_0a_1 \ldots a_n} - 1}{2}}$ and
${\hat{\mu}}_{\frac{\underline{a_0a_1 \ldots a_n} - 1}{2}}$, where,
it has been assumed that all these $2n + (a_1 + a_2 + \ldots a_{n -
1})$ number of random variables are independent and uniformly
distributed on $S_2$. Given the measurement direction $\hat{a} \in
S_2$, Alice calculates her output as
$$\alpha = -\left(\frac{1 +
f_{\frac{\underline{a_0a_1 \ldots a_n} -
1}{2}}}{2}\right)\left[\left(\frac{\frac{\underline{a_0a_1 \ldots
a_n} - 1}{2} + 1}{2}\right) {\rm Sgn}
\left(\hat{a}.{\hat{\lambda}}_{\frac{\underline{a_0a_1 \ldots a_n} -
1}{2}}\right) + \alpha\left(\frac{\underline{a_0a_1 \ldots a_{n -
1}} - 1}{2}\right)\right]$$
\begin{equation}
\label{integeralpha}
\equiv -\alpha\left(\frac{\underline{a_0a_1
\ldots a_n} - 1}{2}\right),
\end{equation} and she sends the $n$ cbits
\begin{equation}
\label{cbitsinteger} c_{k} = {\rm Sgn}
(\hat{a}.{\hat{\lambda}}_k)~ {\rm Sgn} (\hat{a}.{\hat{\mu}}_k),
\end{equation}
to Bob where $k = \frac{\underline{a_0a_1 \ldots a_n} - 1}{2},
\frac{\underline{a_0a_1 \ldots a_{n - 1}} - 1}{2}, \ldots,
\frac{\underline{a_0a_1} - 1}{2}$. After receiving these $n$ cbits
and using his measurement direction $\hat{b} \in S_2$, Bob
calculates his output as
$$\beta =
\left(\frac{1 + f_{\frac{\underline{a_0a_1 \ldots a_n} -
1}{2}}}{2}\right)\left[\left(\frac{\frac{\underline{a_0a_1 \ldots
a_n} - 1}{2} + 1}{2}\right) {\rm Sgn}
\left[\hat{b}.\left({\hat{\lambda}}_{\frac{\underline{a_0a_1 \ldots
a_n} - 1}{2}} + c_{\frac{\underline{a_0a_1 \ldots a_n} -
1}{2}}{\hat{\mu}}_{\frac{\underline{a_0a_1 \ldots a_n} -
1}{2}}\right)\right]\right.$$
\begin{equation}
\label{integerbeta} \left.+ \beta\left(\frac{\underline{a_0a_1
\ldots a_{n - 1}} - 1}{2}\right)\right]
 \equiv \beta\left(\frac{\underline{a_0a_1 \ldots
a_n} - 1}{2}\right).
\end{equation}
Here \begin{equation} \label{fvalue} f_{\frac{\underline{a_0a_1
\ldots a_n} - 1}{2}} = {\rm Sgn}
\left(\hat{z}.{\hat{\nu}}_{\frac{\underline{a_0a_1 \ldots a_n} -
1}{2}} + \frac{\underline{a_0a_1 \ldots a_n}-2}{\underline{a_0a_1
\ldots a_n}}\right).
\end{equation}
Let $L = a_1 + a_2 + \ldots
a_n$ and let $i_1$, $i_2$, $\ldots$, $i_L$ be elements from $\{1, 2,
\ldots, n\}$ such that $i_1 < i_2 < \ldots < i_L$ and $a_{i_1} =
a_{i_2} = \ldots = a_{i_L} = 1$. It is then easy to see that
$$S_{\lambda} = \left\{{\hat{\lambda}}_{\frac{\underline{a_0a_1} -
1}{2}}, {\hat{\lambda}}_{\frac{\underline{a_0a_1a_2} - 1}{2}},
\ldots, {\hat{\lambda}}_{\frac{\underline{a_0a_1 \ldots a_n} -
1}{2}}\right\},$$
$$S_{\mu} = \left\{{\hat{\mu}}_{\frac{\underline{a_0a_1} -
1}{2}}, {\hat{\mu}}_{\frac{\underline{a_0a_1a_2} - 1}{2}}, \ldots,
{\hat{\mu}}_{\frac{\underline{a_0a_1 \ldots a_n} - 1}{2}}\right\},$$
$$S_{\nu} = \left\{{\hat{\nu}}_{\frac{\underline{a_0a_{i_1}} -
1}{2}}, {\hat{\nu}}_{\frac{\underline{a_0a_{i_1}a_{i_2}} - 1}{2}},
\ldots, {\hat{\nu}}_{\frac{\underline{a_0a_{i_1} \ldots a_{i_L}} -
1}{2}}\right\}.$$

\vspace{0.2cm} The way we have defined
$\alpha\left(\frac{\underline{a_0a_1 \ldots a_n} - 1}{2}\right)$ as
well as $\beta\left(\frac{\underline{a_0a_1 \ldots a_n} -
1}{2}\right)$ (see examples (1.1) - (2.4) as well as equations
(\ref{halfintegeralpha}), (\ref{halfintegerbeta}),
(\ref{integeralpha}) and (\ref{integerbeta})), one can show
recursively that
$${\rm Prob} \left(\alpha\left(\frac{\underline{a_0a_1 \ldots a_n} -
1}{2}\right) = j\right) = {\rm Prob}
\left(\beta\left(\frac{\underline{a_0a_1 \ldots a_n} - 1}{2}\right)
= k\right) = \frac{1}{\underline{a_0a_1 \ldots a_n}}$$ for $j, k \in
\{(\underline{a_0a_1 \ldots a_n} - 1)/2, (\underline{a_0a_1 \ldots
a_n} - 3)/2, \ldots, -(\underline{a_0a_1 \ldots a_n} - 3)/2,
-(\underline{a_0a_1 \ldots a_n} - 1)/2\}$ and also
$$\langle \alpha \beta \rangle = \left\langle -\alpha\left(\frac{\underline{a_0a_1 \ldots a_n} -
1}{2}\right) \times \beta\left(\frac{\underline{a_0a_1 \ldots a_n} -
1}{2}\right) \right\rangle$$
$$= -\frac{1}{3} \times
\frac{\underline{a_0a_1 \ldots a_n} - 1}{2} \times
\left(\frac{\underline{a_0a_1 \ldots a_n} - 1}{2} +
1\right)\left(\hat{a}.\hat{b}\right) = \langle \alpha \beta
\rangle_{QM}$$

Thus we see that for any given value of the spin $s$ (integer or
half-integer) for which $2^n - 1 < d = 2s + 1 \le 2^{n + 1} - 1$
(hence $d$ has the binary representation $d = \underline{a_0a_1
\ldots a_n}$ where $a_0$, $a_1$, $\ldots$, $a_n \in \{0, 1\}$ and
$a_0 \ne 0$), Alice and Bob can simulate, in the worst case
scenario,  the measurement correlation in the two spin-$s$ singlet
state $|{\psi}^-_s\rangle$ for performing measurement of arbitrary
spin observables by using only $n = \lceil{\rm log}_2 (s + 1)\rceil$
bits of communication if they {\it a priori} share $2n + (a_1 + a_2
+ \ldots a_n)$ number of independent and uniformly distributed
random variables on $S_2$.

For any maximally entangled state $|{\psi}_{max}\rangle$ of two
spin-$s$ systems, we know that there exists a $(2s + 1) \times (2s +
1)$ unitary matrix $U$ such that $|{\psi}_{max}\rangle = (U \times
I)|{\psi}^-_s\rangle$. Our protocol works equally well for those two
spin-$s$ maximally entangled state $|{\psi}_{max}\rangle$ for each
of which the above-mentioned unitary matrix $U$ induces a rotation
in ${I\!\!\!\!R}^3$, as in those cases, both Alice and Bob can
perform the protocol for the spin-£s£ singlet state
$|{\psi}^-_s\rangle$ for the rotated input vectors $\hat{a}$ and
$\hat{b}$ and, hence, they will achieve their goal.

\section{Conclusion}
Our result provides the amount of classical communication in the
worst case scenario if we consider only measurement of spin
observables on both sides of a two spin-$s$ singlet state for all
the values of $s$ -- just $n = \lceil{\rm log}_2 (s + 1)\rceil$ bits
of communication from Alice to Bob is sufficient. Thus, in our
simulation protocol, the required amount of classical communication
is increased only by one cbit if dimension of the individual spin
system becomes double. In other words, the amount of classical
communication, in our simulation scheme, is equal to the maximum
number of qubit(s) one can accommodate within the Hilbert space
dimension of the individual spin system.

It should be noted that if we consider most general projective
measurements on both the sides of a maximally entangled state of two
qudits, with $d = 2^n$, it is known that (see \cite{brassard99})
Alice would require at least of the order of $2^n$ bits of
communication to be sent to Bob, in the worst case scenario when $n$
is large enough. But for general $d$, ${\rm log}_2 d$ can be shown
to be a lower bound on the average amount of classical communication
that one would require to simulate the maximally entangled
correlation of two qudits considering most general type of
projective measurements \cite{barrett06}. So, in the worst case
scenario, one would require at least ${\rm log}_2 d$ number of bits
of communication for simulating measurement correlation of the
two-qudit maximally entangled state, where the measurement can be
arbitrary but projection type. If one can show that ${\rm log}_2 d$
is again a lower bound for considering measurement of spin
observables only (which we believe to be true), our simulation
scheme will turn out to be optimal.

\vspace{0.3cm} {\textbf{Acknowledgement:}} We thank Guruprasad Kar
and R. Simon for general encouragement. We would like to thank Ali Saif M.Hassan for drawing figure 1 for the present paper.

\end{document}